\documentclass[12pt]{article}
\usepackage{graphicx}
\usepackage{amssymb}
\newcommand{\beq}{\begin{equation}}
\newcommand{\eeq}{\end{equation}}
\newcommand{\beqn}{\begin{eqnarray}}
\newcommand{\eeqn}{\end{eqnarray}}

\newcommand{\al}{\mbox{${\alpha}$}}
\newcommand{\ga}{\mbox{${\gamma}$}}
\newcommand{\ep}{\mbox{${\varepsilon}$}}
\newcommand{\lsim}{\mbox{$<$\hspace{-0.8em}\raisebox{-0.4em}{$\sim$}}}

\begin{document}

\begin{center}
{\Large \bf  Upper limits on electric dipole moments\\ of
$\tau$-lepton, heavy quarks, and $W$-boson}
\end{center}

\vspace{1cm}

\begin{center}
A.G.~Grozin\footnote{A.G.Grozin@inp.nsk.su}, I.B.~Khriplovich\footnote{khriplovich@inp.nsk.su}, and A.S.~Rudenko\footnote{saber\_@inbox.ru}\\
Budker Institute of Nuclear Physics\\
630090 Novosibirsk, Russia,\\
and Novosibirsk University
\end{center}

\vspace{1cm}

\begin{abstract}
We discuss upper limits on the electric dipole moments (EDM) of
the $\tau$-lepton, heavy quarks, and $W$-boson, which follow from
the precision measurements of the electron and neutron EDM.
\end{abstract}

\vspace{1cm}

\section{Introduction}
Strict upper limits on the electric dipole moments of common elementary particles,
electron and proton, were derived from spectroscopic, almost table-top experiments
\cite{bcr,for,ds}. As to the neutron EDM, the best upper limit on it was obtained as
a result of reactor experiments lasting many years \cite{bak} (they say that the
searches for the neutron EDM killed more theories than any other experiment in the
history of physics). And at last, the result for the muon EDM follows from the
measurements at the dedicated muon storage ring \cite{jmb}. These results are
summarized in Table~1.

\begin{center}\footnotesize{
\begin{tabular}{|c|c|c|c|c|} \hline
& & & &  \\
& $e$  & $p$ & $n$ & $\mu$  \\
& & & &  \\ \hline
& & & &  \\
$d/e$, cm & $(0.7\pm 0.7)\times 10^{-27}$ \cite{bcr}&$<0.8\times 10^{-24}$
\cite{for,ds} &$<0.29\times 10^{-25}$ \cite{bak}
&$(0.37\pm 0.34)\times 10^{-18}$ \cite{jmb}\\
& & & &  \\ \hline

\end{tabular}}

\vspace{5mm} Table 1

\end{center}

As to the dipole moments of the $\tau$-lepton and heavy quarks, upper limits on them
have been obtained up to now from the analysis of high-energy experiments.

The approach pursued here is based on the precision results \cite{bcr,bak}. We
establish upper limits on the dipole moments of the $\tau$-lepton and heavy quarks
through the analysis of their possible contributions to the electron EDM. This is a
clean theoretical problem for the $\tau$-lepton. For heavy quarks and $W$-boson
these our results are of rather qualitative nature. Additional upper limits on the
dipole moments of heavy quarks, and $W$-boson, also qualitative ones, are derived by
the analysis of their possible contributions to the neutron EDM.

\section{Dipole moment of $\tau$-lepton and electron  EDM}
\vspace{5mm}
\begin{figure}[ht]
\center
\begin{tabular}{c c c c}
\includegraphics{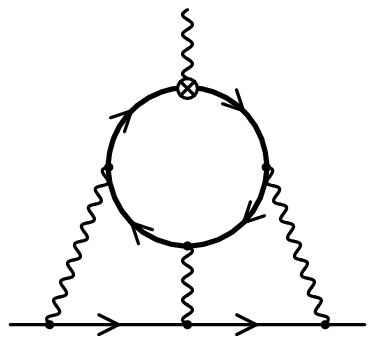} &
\includegraphics{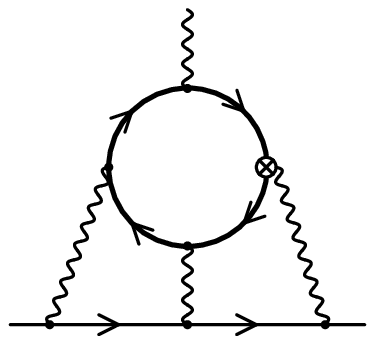} &
\includegraphics{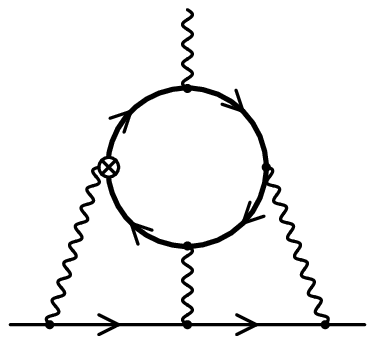} &
\includegraphics{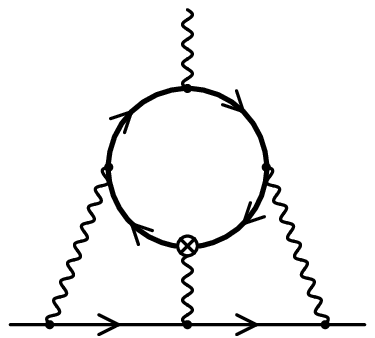} \\
a & b & c & d

\end{tabular}
\vspace{3 mm} Figure 1

\end{figure}

We start with the analysis of the contribution of the $\tau$ EDM $d_\tau$ to the
electron dipole moment $d_e$. This contribution is described by the diagrams of the
type presented in Figs. 1a,b,c,d. Here the loop is formed by the $\tau$ line, and
the lower solid line is the electron one. The upper wavy line corresponds to the
external electric field. The crossed vertices refer to the electromagnetic
interaction of the $\tau$ EDM \beq\label{d1/2} L_{\tau}^{edm} = \,-\,
\frac{1}{2}\,d_{\tau}\, \overline{\tau}\ga_5 \sigma_{\mu\nu}\tau\, F_{\mu\nu}\,=
\,i\, \frac{1}{2}\,d_{\tau}\, \overline{\tau} \sigma_{\mu\nu}\tau\,
\tilde{F}_{\mu\nu}\,; \quad \tilde{F}_{\mu\nu}\, = \, \frac{1}{2}\,
\ep_{\mu\nu\alpha\beta}F_{\alpha\beta}\,. \eeq Of course, all six permutations of
the electromagnetic vertices on the electron line should be considered. The
contributions of diagrams 1b and 1c are equal.

This problem is similar to that of the contribution by the light-by-light scattering
via muon loop to the electron magnetic moment~\cite{rem}. The general structure of
the resulting contribution to the electron EDM is rather obvious (to the leading
order in $m_e/m_\tau$): \beq\label{est} \Delta d_e =\, a\, \frac{m_e}{m_\tau}
\left(\frac{\al}{\pi}\right)^3 d_\tau\,, \eeq where $a$ is some numerical factor
(hopefully, on the order of unity). The factor $m_e$ originates from the necessary
helicity-flip on the electron line; then $1/m_\tau$ is dictated by dimensional
arguments.

Diagram 1a corresponds to the matrix element
${<}e|\bar{\tau}\sigma^{\mu\nu}\tau|e{>}=C\bar{u}\sigma^{\mu\nu}u$. We use
dimensional regularization with $d=4-2\ep$ dimensions and the method of regions (see
the textbook~\cite{Smirnov}). Only the region where all three loops are hard (loop
momenta $\sim m_\tau$) contributes to the leading power term~(\ref{est}); therefore,
there are no logarithms $\ln(m_\tau/m_e)$ (contributions of regions with 1 or 2 hard
loops are suppressed by an extra factor $(m_e/m_\tau)^2$). In this hard region, the
problem reduces to 3-loop vacuum integrals with a single mass $m_\tau$ belonging to
the simpler topology $B_M$~\cite{Broadhurst}. We perform the calculation in
arbitrary covariant gauge, and use the \textsc{Reduce} package
\textsc{Recursor}~\cite{Broadhurst} to reduce scalar integrals to two master
integrals. Gauge-dependent terms cancel, and we get
\begin{equation}
C = \frac{m_e}{m_\tau} \frac{e^6 m_\tau^{-6\ep}}{(4\pi)^{3d/2}} \Gamma^3(\ep)
\frac{8}{d(d-1)(d-5)} \left[ - 2 \frac{2d^2-21d+61}{d-5} +
\frac{d^4-9d^3+8d^2+84d-126}{2d-9} R \right]\,, \label{C}
\end{equation}
where
\begin{equation}
R =
\frac{\Gamma(1-\ep) \Gamma^2(1+2\ep) \Gamma(1+3\ep)}%
{\Gamma^2(1+\ep) \Gamma(1+4\ep)} = 1 + 8 \zeta(3) \ep^3 + \cdots\,, \label{R}
\end{equation}
and $\zeta$ is the Riemann $\zeta$-function. All divergences cancel, and we arrive
at the finite contribution to $a$: \beq\label{fin1} a_1 = \frac{3}{2}\,\zeta(3) -
\frac{19}{12}\ . \eeq

In order to calculate the contribution of Fig.~1b,c,d, we expand the corresponding
initial expressions in the external photon momentum $q$ up to the linear term. The
EDM vertex contains $\ep^{\mu\nu\alpha\beta}$; we put this factor aside, and
calculate tensor diagrams with four indices. After summing all diagrams, the result
is finite; now we can set $\ep\to 0$, and multiply by $\ep^{\mu\nu\alpha\beta}$
(cf.~\cite{Larin}). The result has the structure of a tree diagram with the electron
EDM vertex $\ep^{\mu\nu\alpha\beta}q_\nu\sigma_{\alpha\beta}$. The gauge-dependent
terms in it cancel (exactly in $d$), as well as the divergences. This contribution
to $a$ is \beq\label{fin2} a_2 = \frac{9}{4}\,\zeta(3) - 1\,. \eeq

As an additional check of our programs, we have reproduced the leading power term in
the contribution to the electron magnetic moment originating from the light-by-light
scattering via the muon loop (formula~(4) in~\cite{rem}).

The final result for the numerical coefficient is \beq\label{fin} a= a_1 + a_2 =
\frac{15}{4}\, \zeta(3) - \frac{31}{12} = 1.924\,. \eeq With this value of $a$, the
discussed contribution to the electron EDM is \beq \Delta d_e = 6.9 \times
10^{-12}\, d_\tau. \eeq Combining this result with the experimental one \cite{bcr}
(see Table 1) for the electron EDM, we arrive at \beq\label{res} d_\tau/e = (1 \pm
1)\times 10^{-16}\,\mathrm{cm}\,. \eeq

In fact, the results (\ref{fin1}) and (\ref{fin2}) refer to somewhat different
regions of incoming momenta. For (\ref{fin2}) all the three momenta are hard, on the
order of magnitude about $m_\tau$, but for (\ref{fin1}) only two of them belong to
this region, and the third one, that of the outer photon, is soft, of vanishing
momentum. Still, one may expect that the effective EDM interaction is formed at
momenta much higher than $m_\tau$, so that this difference is not of much
importance. Besides, the contribution of diagram 1a is anyway numerically small.
Thus, result (\ref{fin1}) is valid at least for all momenta about $m_\tau \sim 1$ --
2 GeV.

The upper limits on the $\tau$ EDM derived from the accelerator experiments
\cite{Aguila, Inami, Escribano, br} belong to the interval of $10^{-16} - 10^{-17}$
$e\cdot$cm, so that our result (\ref{res}) formally does not improve them. However,
all those accelerator data refer to much larger typical momenta of the photon, from
10 to 200 GeV.

\section{Dipole moments of heavy quarks and electron EDM}

The information on the dipole moments $d_q$ of heavy quarks can be obtained from the
diagrams analogous to those presented in Figures 1, but with quarks in the fermion
loops, instead of the $\tau$-lepton. We do not consider the gluon corrections to the
quark loops, and confine to the simple estimate for them, following from the analogy
between the two problems (see (\ref{est})): \beq\label{esq} \Delta d_e = a_q\,3\,
Q^3\, \frac{m_e}{m_q} \left(\frac{\alpha}{\pi}\right)^3 d_q\,; \eeq here $m_q$ is
the quark mass, $Q$ is the quark charge in the units of $e$ ($Q_{b} = -1/3$,
$Q_{c,t} = 2/3$). As to the overall numerical factor here, we put $a_q \simeq 1$.
The corresponding estimates are straightforward. Here and below we assume the
following values for the quark masses:
\begin{center} $m_c = 1.25$
GeV; \quad $m_b = 4.5$ GeV; \quad $m_t = 175$ GeV.\end{center} Thus obtained upper
limits are: \beq\label{lim1} d_c/e \; \lsim \; 3 \times 10^{-16}\; {\rm cm}; \quad
d_b/e \; \lsim \;7 \times 10^{-15}\; {\rm cm};\quad d_t/e \;\lsim \; 4 \times
10^{-14}\; {\rm cm}\,. \eeq

\section{Dipole moments of heavy quarks and neutron EDM}

The dipole moment of a heavy quark $Q$ generates the EDM of a light quark $q$ via
the following diagram \cite{cc}:
\begin{figure}[h]
\begin{center}
\includegraphics[scale=1.2]{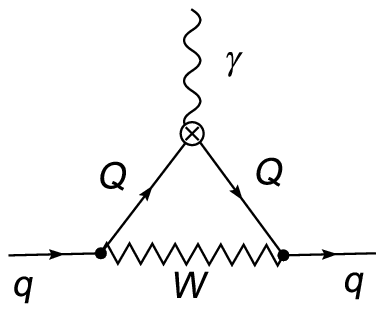}

\vspace{3mm} Figure 2
\end{center}
\end{figure}

The scattering amplitudes and corresponding contributions to the dipole moments of
light quarks are: \beq\label{QqM} M_1 = \frac{\alpha}{16\pi
\sin^2\theta_w}\,|V_{Qq}|^2\,\frac{m_q \,m_Q}{m_W^2}
\left(\ln\frac{\Lambda^2}{m_W^2}- 2\right)
\frac{1}{2}\,d_Q\bar{u}\gamma_5\sigma_{\mu\nu}u\,F_{\mu\nu}\,, \eeq \beq\label{Qq}
\Delta d_q = \frac{\alpha}{16\pi \sin^2\theta_w}\,|V_{Qq}|^2\,\frac{m_q
\,m_Q}{m_W^2} \left(\ln\frac{\Lambda^2}{m_W^2}- 2\right)d_Q, \eeq for
\[
Q = c,\;\; q = d \quad \mathrm{or} \quad Q = b,\;\; q = u;
\]
and \beq\label{tdM} M_2 = \frac{\alpha}{16\pi\sin^2\theta_w}\,|V_{td}|^2\,\frac{m_t
m_d}{m_W^2} \left(\ln\frac{\Lambda^2}{m_t^2}- 2\right)
\frac{1}{2}\,d_t\,\bar{u}\gamma_5\sigma_{\mu\nu}u\,F_{\mu\nu}\,, \eeq \beq\label{td}
\Delta d_d = \frac{\alpha}{16\pi \sin^2\theta_w}\,|V_{td}|^2\,\frac{m_t m_d}{m_W^2}
\left(\ln\frac{\Lambda^2}{m_t^2}- 2\right)d_t, \eeq for
\[
Q = t,\;\; q = d.
\]
Here $V_{Qq}$ are the corresponding coefficients of the Kobayashi-Maskawa matrix;
$\theta_w$ is the Weinberg angle, $\sin^2 \theta_w = 0.23$. The structures of
relations (\ref{Qq}) and (\ref{td}) are different since $m_{c,\,b} \ll m_W$, while
$m_t
> m_W$. In fact, to simplify final expressions, we confine to the
leading order in $1/m_W^2$ in (\ref{QqM}), (\ref{Qq}), and even assume that $m_t \gg
m_W$ in (\ref{tdM}), (\ref{td}). In both cases the diagrams are logarithmically
divergent, so that we have to introduce a cut-off $\Lambda$ at high momenta.

This divergence is caused by the term $k_\mu k_\nu/m^2_W$ in the numerator of the
$W$ Green function
\[
\frac{-\delta_{\mu\nu} + k_{\mu} k_{\nu}/m_W^2}{k^2 - m_W^2}\,.
\]
The mentioned term $k_\mu k_\nu/m^2_W$ was omitted in \cite{cc}. Therefore,
relations (\ref{Qq}) and (\ref{td}) differ essentially from the corresponding
results of \cite{cc}.

In our estimates here, we assume that both $\ln(\Lambda^2/m_W^2)- 2$ and
$\ln(\Lambda^2/m_t^2)- 2$ are on the order of unity. As to the light quark dipole
moments, we assume (in the spirit of the constituent quark model) that they are on
the same order of magnitude as the neutron EDM:
\[
d_{u,\,d} \sim d_n.
\]
Of course, both these assumptions make the corresponding estimates less definite
than those based on electron EDM. The results of our estimates are presented in the
last line of Table 2. In its previous line we repeat for comparison the data already
given in (\ref{lim1}). In both sets of estimates the strong interaction of quarks is
neglected.

\begin{center}
\begin{tabular}{|c|c|c|c|} \hline
 & & &  \\
 & $c$ & $b$ & $t$  \\
  & & &  \\
\hline
 & & &  \\
$m_Q$, GeV  & 1.25 &  4.5 & 175 \\
 & & &  \\
\hline
& & &  \\
$d_Q/e$, cm & $\lsim\; 3 \times 10^{-16}$& $\lsim\; 7 \times 10^{-15}$&
$\lsim\; 4 \times 10^{-14}$\\
derived with $d_e$ & & &  \\ \hline
 & & &  \\
$d_Q/e$, cm & $\lsim \; 10^{-15}$ & $\lsim \; 2 \times 10^{-12}$&
$\lsim\; 5 \times 10^{-15}$\\
derived with $d_n$ & & &  \\ \hline

\end{tabular}

\vspace{5mm} Table 2. Limits on quark dipole moments from electron and neutron EDMs

\end{center}

It is only natural that for the most heavy $t$-quark the bound from $d_n$ dominates.

The bounds on the dipole moments of $c$- and $b$-quarks derived from the high-energy
experiments \cite{br, Escribano2} are on the order of $10^{-17}$ $e\cdot$cm, so that
they are more strict than the limits given in Table~2. However, as noted above
already, the high-energy data refer to different, much larger typical momenta of the
photon.

In conclusion of this section, we mention the investigation of the radiative decay
$b \to s\ga$ as a probe of the $t$\,-quark EDM \cite{rizzo}. The discussed
contribution to the decay amplitude is described by the diagram in Fig.~3.
\begin{figure}[h]
\begin{center}
\includegraphics[scale=1.2]{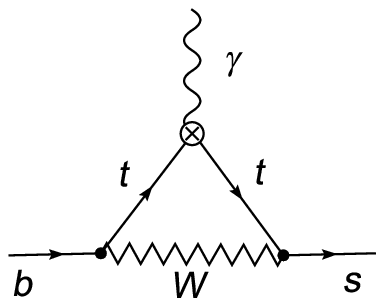}

\vspace{3mm} Figure 3
\end{center}
\end{figure}
This diagram is also logarithmically divergent. Its contribution can be easily
obtained as follows. In the limit $m_b \to m_s$, the structure of discussed matrix
element should coincide with that of (\ref{tdM}) (up to the obvious changes $m_d \to
m_{b(s)}$ and $|V_{td}|^2 \to V^*_{ts} V_{tb}$). On the other hand, in the limit
$m_s \to 0$, the outgoing $s$-quark should be left-handed. Therefore, the matrix
element of the $b \to s\ga$ decay (with the internal $t$-quark) should look as
follows:
\[
M(b \to s\ga)= \frac{\alpha}{64\pi\sin^2\theta_w}\,V^*_{ts}
V_{tb}\,\frac{m_t}{m_W^2} \left(\ln\frac{\Lambda^2}{m_t^2}- 2\right) d_t\,\bar{u}_s
\Bigg[- m_b(1-\gamma_5) +m_s(1+\gamma_5)\Bigg]\sigma_{\mu\nu}u_b\,F_{\mu\nu}\,
\]
\beq \simeq -\,\frac{\alpha}{64\pi\sin^2\theta_w}\,V^*_{ts} V_{tb}\,\frac{m_b
m_t}{m_W^2} \left(\ln\frac{\Lambda^2}{m_t^2}- 2\right) d_t\,\bar{u}_s(1-\gamma_5)\,
\sigma_{\mu\nu}u_b\,F_{\mu\nu}\,. \eeq The result presented in \cite{rizzo} is \beq
d_t/e \; \lsim \; 10^{-16}\; \rm{cm}. \eeq

\section{Electric dipole moment of $W$-boson}

\begin{figure}[h]
\center
\includegraphics[scale=0.9]{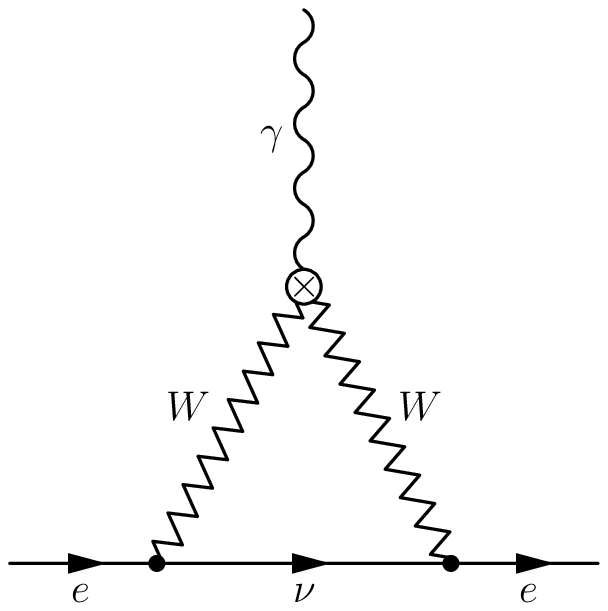}

\vspace{3mm} Figure 4
\end{figure}

One more contribution to the electron and neutron dipole moments can be given by the
EDM $d_W$ of $W$-boson. This effect was pointed out and investigated long ago
\cite{ss,mq}.

It is convenient to start the discussion with the electron EDM. The effect is
described by diagram presented in Fig.~4. Here as well the crossed vertex refers to
the electromagnetic interaction of the $W$-boson EDM (it is obvious from the diagram
that here it is the dipole moment of $W^-$). In this case the EDM interaction is
described by the Lagrangian \beq\label{d1} L_W^{edm} = 2m_W i d_W
\tilde{F}_{\alpha\beta} W^\dag_{\alpha} W_\beta\,. \eeq The corresponding matrix
element is
\[
M = \frac{\pi \alpha}{\sin^2\theta_w}\;d_W\,m_W \int\frac{d^4 k}{(2\pi)^4}
\,\bar{u}_e(p)\gamma_\mu (1+\ga_5)\,\frac{\hat{k}}{k^2}\,\gamma_\nu
(1+\ga_5)u_e(p)\times
\]
\beq\label{mel} \times\,\frac{1}{[(k-p)^2 -
m_W^2]^2}\,\left\{\tilde{F}_{\mu\nu}-\,\frac{1}{m_W^2}\left[(k-p)_\mu (k-p)_\alpha
\tilde{F}_{\alpha\nu} + \tilde{F}_{\mu\alpha}(k-p)_\alpha(k-p)_\nu\right]\right\}.
\eeq With straightforward, though rather tedious calculations (somewhat simplified
by employing the density matrix of polarized fermion), one arrives at the following
result for the contribution of $W$-boson EDM to electron dipole moment:
\beq\label{del} \Delta d_e = \,\frac{\alpha}{8\pi \sin^2
\theta_w}\,\frac{m_e}{m_W}\,\ln\frac{\Lambda^2}{m_W^2}\,d_W. \eeq Here $\Lambda$ is
the cut-off parameter for the logarithmically divergent integral over virtual
momenta in the loop. Putting (perhaps, quite conservatively) $\ln(\Lambda^2/m_W^2)
\simeq 1$, one obtains with the experimental upper limit on the electron EDM
\cite{bcr} (see Table 1), the following bound on the dipole moment of $W$-boson:
\beq\label{We} d_W/e \; \lsim\; 2 \times 10^{-19} \; \rm{cm}\,. \eeq

In the case of the $W$-boson contribution to the neutron EDM, our line of reasoning
somewhat differs from that of \cite{ss,mq}. We note first of all that the electron
mass $m_e$ does not enter explicitly matrix element (\ref{mel}). It arises in the
result (\ref{del}) only as the mass of an external fermion, via the Dirac equation
$\hat{p}\,u = m_e u$. Therefore, there are all the reasons to expect that the
contribution of $d_W$ to the neutron EDM will be proportional to the neutron mass
$m_n$, i.e. enhanced as compared to (\ref{del}) by three orders of magnitude. In
this case, the forward scattering amplitude of the virtual $W$-boson can be written
in a general form as follows\footnote{Compare with the corresponding structure
$\bar{u}_e(p)\gamma_\mu (1+\ga_5)\,(\hat{k}/k^2)\,\gamma_\nu (1+\ga_5)u_e(p)$ in
formula (\ref{mel}) for electron.}: \beq\label{amp} \bar{u}_n(p)\gamma_\mu
(1+\ga_5)\,\left[\hat{k}\,g(k^2)+ \,\hat{p}\,h(k^2)\right]\gamma_\nu
(1+\ga_5)u_n(p)\,; \eeq
 here $k$ is the total momentum of intermediate hadronic states. Of course, the invariant functions $g$
 and $h$ depend in fact not only on $k^2$, but on $(kp)$ as well.
 However, in our case $k^2 \sim m_W^2 \gg (kp) \sim m_n m_W$, so that the dependence on $(kp)$ can be safely
 ignored. By the analogous reason, in the usual case of the deep
 inelastic neutrino scattering, the structure with $\hat{p}$ in
 the corresponding amplitude is also omitted. In
 the present case, however, we should keep in amplitude (\ref{amp}) $\hat{p}$, in addition to the common
$\hat{k}$, since after integrating over
 $d^4 k$ both structures give comparable contributions to the result.

 At last, the usual dimensional and scaling
 arguments dictate that asymptotically, for $k^2 \sim m_W^2$, both functions $g$ and $h$ behave as follows:
 \[
g(k^2) = \frac{g_0}{k^2}, \quad h(k^2) = \frac{h_0}{k^2}\,.
 \]
In particular, one can neglect the gluon corrections in these functions. Without any
additional parameters, it is natural to assume that $g_0,\, h_0 \sim 1$.

Now, the same calculations as those in the case of electron EDM, result in the
following expression for the discussed contribution to the neutron dipole moment:
\beq\label{dn} \Delta d_n = \,\frac{\alpha}{8\pi \sin^2
\theta_w}\,\frac{m_n}{m_W}\left[g_0 \ln\frac{\Lambda^2}{m_W^2}\, +
h_0\left(\ln\frac{\Lambda^2}{m_W^2} + 1\right)\right]d_W. \eeq For numerical
estimate we put \[g_0 \ln\frac{\Lambda^2}{m_W^2}\, +
h_0\left(\ln\frac{\Lambda^2}{m_W^2} + 1\right) \sim 1\,,\]  so that
\[
\Delta d_n \sim \,\frac{\alpha}{8\pi \sin^2 \theta_w}\,\frac{m_n}{m_W}\,d_W\,\approx
\,\frac{\alpha}{2\pi}\,\frac{m_n}{m_W}\,d_W\,.
\]
Then, with the result of \cite{bak} for the neutron EDM (see Table~1), we arrive at
the following quite strict upper limit on the $W$-boson dipole moment:
\beq\label{Wn} d_W/e \; \lsim\; 2 \times 10^{-21} \; \rm{cm}\,. \eeq

\bigskip

{\bf Acknowledgements.} The work was supported in part by the Russian Foundation for
Basic Research through Grant No. 08-02-00960-a.


\begin{thebibliography}{}

\bibitem{bcr}
B.C.~Regan {\em et al}, Phys.\ Rev.\ Lett.\ \textbf{88}, 071805 (2002).

\bibitem{for}
W.C.~Griffith {\em et al}, arXiv:0901.2328.

\bibitem{ds}
V.F.~Dmitriev, R.A.~Sen'kov, Phys.\ Rev.\ Lett.\ \textbf{91}, 212303 (2003).

\bibitem{bak}
C.A.~Baker {\em et al}, Phys.\ Rev.\ Lett.\ \textbf{97}, 131801 (2006).

\bibitem{jmb}
J.M.~Bailey {\em et al}, J.\ Phys.\ G \textbf{4}, 345 (1978).

\bibitem{rem}
S.~Laporta, E.~Remiddi, Phys.\ Lett.\ B \textbf{301}, 440 (1993).

\bibitem{Smirnov}
V.A.~Smirnov, \textit{Applied asymptotic expansions in momenta and masses}, Springer
Tracts in Modern Physics \textbf{177}, Springer (2002), Chapter~5.

\bibitem{Broadhurst}
D.J.~Broadhurst, Z.\ Phys.\ C \textbf{54}, 599 (1992).

\bibitem{Larin}
S.A.~Larin,
%in \textit{Quarks-92},
%ed.\ D.Yu.~Grigoriev, V.A.~Matveev, V.A.~Rubakov, P.G.~Tinyakov,
%World Scientific (1993), p.~201;
Phys.\ Lett.\ B \textbf{303}, 113 (1993).

\bibitem{Aguila}
F.~del~Aguila, M.~Sher, Phys. Lett. B \textbf{252}, 116 (1990).

\bibitem{Inami}
K.~Inami {\em et al}.\ (Belle Collaboration), Phys.\ Lett.\ B \textbf{551}, 16
(2003).

\bibitem{Escribano}
R.~Escribano, E.~Mass\'o, Phys. Lett. B \textbf{395}, 369 (1997).

\bibitem{br}
A.E.~Blinov, A.S.~Rudenko,  arXiv:0811.2380.

\bibitem{cc}
A.~Cordero-Cid, J.M.~Hernandez, G.~Tavares-Velasco, J.J.~Toscano, J. Phys. G
\textbf{35}, 025004 (2008).

\bibitem{Escribano2}
R.~Escribano, E.~Mass\'o, Nucl. Phys. B \textbf{429}, 19 (1994).

\bibitem{rizzo}
J.L.~Hewett, T.M.~Rizzo, Phys.\ Rev.\ D \textbf{49}, 319 (1994).

\bibitem{ss} F.~Salzman, G.~Salzman, Phys. Lett. \textbf{15}, 91 (1965);
Nuovo Cimento A \textbf{41}, 443 (1966).

\bibitem{mq} F.J.~Marciano, A.~Queijeiro, Phys. Rev. D \textbf{33}, 3449 (1986).

\end{thebibliography}
\end{document}